# Correlation between extrinsic electroresistance and magnetoresistance in fine-grained $La_{0.7}Ca_{0.3}MnO_3$

Parukuttyamma Sujatha Devi, Abhoy Kumar,[†] and Dipten Bhattacharya[*]

*Nanostructured Materials Division, Central Glass and Ceramic Research Institute, CSIR, Kolkata 700032, India*

Shilpi Karmakar[§] and Bijoy Krishna Chaudhuri

*Department of Solid State Physics, Indian Association for the Cultivation of Science, Kolkata 700032, India*

We report our observation of a correlation between the extrinsic electroresistance (EER) and magnetoresistance (EMR) via grain size in fine-grained $La_{0.7}Ca_{0.3}MnO_3$. The nature of dependence of EER and EMR on grain size (~0.2-1.0 μm) indicates that for finer grains with low-resistive boundaries both of them follow similar trend whereas they differ for coarser grains with high-resistive boundaries. This could be due to a crossover in the mechanism of charge transport across the grain boundaries – from spin-dependent scattering process to spin-polarized tunneling one – as a function of grain size.

PACS Nos. 75.47.-m; 75.47.Lx

___________________________
[*]Corresponding author; e-mail: dipten@cgcri.res.in
[†]Present address: Defense Metallurgical Research Laboratory, Hyderabad 500058, India
[§]Currently at National Nanotechnology Laboratory, 73100 Lecce, Italy

# 1. Introduction

It has been reported first in 1997 that the colossal magnetoresistive manganites not only exhibit large magnetoresistance (MR) but also a sizable electroresistance (ER).[1] Since then, the ER has been studied in a series of manganites both under the field-effect and current driven conditions.[2,3] However, an unsolved issue is whether the origin of both the effects is same or different. In single crystals of doped manganites[2-6] (e.g., in $La_{0.7}Ca_{0.3}MnO_3$, $Nd_{0.7}Sr_{0.3}MnO_3$, $La_{0.5}Ca_{0.5}MnO_3$ etc.), the ER and MR are found to be complimentary to each other: ER is high at higher temperature (above $T_{MI}$; $T_{MI}$: metal-insulator transition point) and MR is high at lower temperature (below $T_{MI}$). It has been argued that the boundary between the two segregated phases – ferromagnetic metallic (FMM) and charge-ordered insulating (COI) – moves *under electric field* establishing connectivity among the FMM islands. This, in turn, gives rise to charge carrier type (p or n) dependent ER yet no shift in transition temperature ($T_{MI}$). On the other hand, the volume fraction of the FMM phase grows *under magnetic field* at the expense of that of COI phase which gives rise to both MR and upward shift in $T_{MI}$. Thus, a picture appears to be emerging to understand the correlation between ER and MR in the *continuum system*. Could there be any correlation or anti-correlation between extrinsic ER (EER) and extrinsic MR (EMR) in *granular systems*[7] too? This question is important as both electric and magnetic fields influence the charge carrier transport across the grain boundaries enormously. It has already been recognized that EMR in the granular systems is more useful than the intrinsic MR for the application of these compounds as magnetic read heads and other magnetoresistive sensors. Discovery of a correlation between EER and EMR will help in efficient tuning of the overall field dependent effects in the granular systems. Simultaneous application of the electric and magnetic fields and variation of one by keeping the other fixed will allow more leverage in controlling the field-dependent drop in electrical resistivity across a certain field-temperature zone. While the electric field drives dielectric breakdown of the insulating grain boundary, the magnetic field moves or rotates the magnetic domains within a grain which, in turn, influences the charge transport across the grain boundary. Moreover, the charge transport across the grain boundaries follows different mechanisms – spin polarized tunneling (SPT) or spin-dependent scattering (SDS).[8,9] In the case of charge transport via SPT, majority carriers tunnel through the grain boundary barrier and the EMR at any given temperature (below the Curie point of the grains) and magnetic field depends on intergrain exchange J, spin polarization P and magnetization of the grains with respect to the saturation magnetization. The grain boundary thickness (d) should be small as tunneling current $I \sim \exp(-d)$ while the resistance should be very high. In cases where these conditions are not fulfilled, the charge transport might follow the normal process of spin-dependent scattering where spin alignment at the grain boundaries determines the transport. Non-parallel spin alignment will scatter the charges. Since with the increase in grain size, the grain boundary characteristics such as resistance, area, thickness etc vary, there might be a correlation between EER and EMR guided by the mechanism of charge transport – SPT or SDS.

In this paper, we report that there, indeed, exists an interesting correlation among EER, EMR, grain boundary resistance, and grain size in the fine-grained $La_{0.7}Ca_{0.3}MnO_3$ (LCMO) system. Finer grains with low grain boundary resistance exhibit high EER and EMR while both of them decrease with increasing grain size and grain boundary resistance. Beyond a certain grain size, EMR exhibits an upturn while EER decreases monotonically without any change in pattern.

## 2. Experiments

The fine-grained granular LCMO samples with grain size ~0.2-1.0 μm have been prepared by controlled heat treatment of nanoscale powder (8-18 nm) at lower temperature (1100°C). The nanoparticles of LCMO have been prepared by two techniques: (i) simple solution chemistry where mixed metal nitrate solutions are allowed to undergo controlled yet self-propagating combustion within suitable fuel-oxidant medium leading to the formation of nanosized particles and (ii) where such solutions are sprayed from a spray pyrolyzer during combustion.[10] The crucial parameters for controlling the particle and grain sizes are (i) the route followed for synthesis of precursor powder (spray pyrolysis yields finer particles) and (ii) the heat treatment time. The powder has been characterized by x-ray diffraction and transmission electron microscopy studies. The grain morphology of the sintered pellets has been studied by scanning electron microscopy. The electrical measurements were done in standard four-probe configuration with silver electrodes. A magnetic field of ~0-15 kOe has been used for measuring the magnetoresistance of the samples. The electric resistance under zero and a finite magnetic field (≤15 kOe) was measured across a temperature range ~77-300 K while the current-voltage characteristics under zero magnetic fields were studied at ~77 K only. The applied current was swept from zero to the maximum in forward bias and then reduced from maximum to zero. It was then swept from zero to the maximum in reverse bias, and finally reduced from maximum in reverse bias to zero. The step size of increase and decrease of the current was ~1 mA. At each step the corresponding voltage was measured. We have recorded the current-voltage loops using different sweep rate. We have also used pulsed dc current with pulse width at 'on' state ~4s, 'off' state ~5s, and height ~1 mA for checking whether Joule heating of the sample due to steady dc current is the major source of electroresistnce or not.

## 3. Results and Discussion

The x-ray diffraction (XRD) patterns of the powder are shown in Fig. 1. The crystallographic structure is found to be either pseudocubic or rhombohedral with hexagonal unit cell. The Rietveld analysis of XRD peak profiles yields the average crystallite size (varying within 8-18 nm), the lattice parameters, and the microstrain of the particles. The crystallite size obtained corroborates the observation made in transmission electron microscopy of the powder (data not shown here). The average particle size too was found to be within 8-18 nm. There is slight variation in the lattice parameter from sample to sample. The grain morphology of the sintered pellets evolves systematically: from a highly porous structure consisting of a network of finer grains to a denser structure with higher grain size (Fig. 1b,c). The grain morphology – aspect ratio, grain size, connectivity etc – has been thoroughly studied using scanning electron microscopy coupled with image analyzer software Image-J. The grain size distribution histograms in representative cases are shown in the insets of Figs. 1b,c. The histograms clearly show the difference in average grain size in two different samples. In Table-I we provide the details such as average particle size of the calcined powder, average grain size of the sintered pellets as estimated from the SEM pictures by the software Image J, and the lattice parameters. The density of the sintered pellets varies within 80-90% of the theoretical density for $La_{0.7}Ca_{0.3}MnO_3$. The composition for the pellets has been verified by inductively coupled plasma spectroscopic measurement.

Figures 2 and 3 summarize our observations on the correlation among grain size, $T_{MI}$, grain boundary resistance with respect to the grain resistance, EER, threshold current ($I_{th}$), and EMR. The EER is estimated from $[R(I<I_{th}) - R(I>I_{th})]/R(I<I_{th})$ and the MR is given by $[R(0) - R(H)]/R(0)$, where $I$ is the applied current and $H$ is the applied magnetic field (~15.0 kOe). In order to extract the genuine EMR data at any particular temperature we used our overall MR data at ~15 kOe and subtract from overall MR the intrinsic MR of a single crystal of $La_{0.7}Ca_{0.3}MnO_3$ at ~15 kOe and at that specific temperature.[11] The intrinsic MR at well below the Curie point is small and is nearly temperature independent. We used the extracted EMR values thus for comparing between EER and EMR in granular $La_{0.7}Ca_{0.3}MnO_3$. The EER does not result from sample heating effect due to high current as we did the measurements at ~77 K where all the samples depict metallic behavior. Heating effect would have given rise to increase in resistivity in this regime and not negative differential resistance (NDR). The sample was immersed in liquid nitrogen bath which helps in reducing the heating of the sample due to current flow via a process of heat dissipation. This has been observed by others as well.[12] Moreover, in order to check whether Joule heating due to current flow is solely responsible for the NDR, we have repeated the tracing of current-voltage loop at ~77 K by employing a pulsed dc current. We have observed similar NDR effect with pulsed current as well. In fact, a platinum resistor attached to the surface of the sample recorded a temperature rise of ~5 K due to high current (~1.0 A). The electrode-sample contact resistance, measured by comparing the data obtained under three-probe (where one voltage and current probe are joined) and four-probe configuration, was found to be varying within ~30-85 m$\Omega$/mm$^2$ (junction area $\approx$17 mm$^2$) over a temperature range 77-300 K. This is an accepted way of finding out the electrode-sample contact resistance and the junction resistance for silver electrodes is quite small.[13] Therefore, even though heating does take place because of current flow, it is not entirely responsible for the large electroresistance observed. The electroresistance results, primarily, from dielectric breakdown of the grain boundaries – a field driven effect.

In the insets of Figs. 2 and 3, we show the zero-field resistivity ($\rho$) versus temperature ($T$) plots, current-voltage characteristics at ~77 K observed under dc current, EMR versus temperature plots, and overall MR versus magnetic field (H) at a specific temperature (~87 K). The $T_{MI}$, noted from the $\rho$-$T$ plots, expectedly decreases with the decrease in grain size (Fig. 2). This is because of increasing lattice defects due to creation of oxygen vacancies in finer grains and consequent disorder at the grain-grain interface. Enhanced defect concentration and disorder destroys long range ferromagnetic order and hence decreases the $T_{MI}$.[14] The metallic part of the zero-field $\rho$-$T$ patterns has been fitted with the empirical expression[15] $\rho = \rho_1 + \rho_2 T^{2.5}$, where $\rho_1$ is the resistivity due to static imperfections, grain boundaries etc and $\rho_2$ accounts for the temperature-dependent part of the overall resistivity. The raw data together with the fitted lines for a few representative samples are shown in Fig. 4. The fitting yields the ratio $\rho_1/\rho_2$ which measures the grain boundary resistance with respect to the grain resistance. This ratio, surprisingly, is found to increase initially with the increase in grain size and then beyond a grain size ~0.6 μm it drops (Fig. 2). *This is a counterintuitive result.* The finer grains are expected to contain enhanced defect states at the surface and hence one should expect $\rho_1/\rho_2$ to increase and $T_{MI}$ to decrease with the decrease in grain size. Here, instead, we observe that in the finer grain size range (i.e., ≤0.6 μm), both the $\rho_1/\rho_2$ and $T_{MI}$ decrease with the decrease in grain size. We explain this apparent contradiction in the following way. We invoke a model of coexistence of FMM nanoshorts with surface defects at the grain boundaries. The nanoshorts are nanosized FMM islands which form at the grain boundaries of finer grains due to phase segregation. The nanoshorts reduce the overall grain boundary resistance in the fine-grained samples and give rise to smaller $\rho_1/\rho_2$ ratio. However, the percolating chain comprising of FMM nanoshorts at the grain boundaries and nanosized FMM islands within the grains forms only at a lower temperature than the bulk sample $T_{MI}$ and hence $T_{MI}$ drops with the drop in grain size. This could be because of influence of enhanced surface scattering at the grain-grain interface and an effective pinning of the FMM shorts and islands preventing percolating chain to form as a consequence. Only at a lower temperature, the chain forms and one observes an insulator-metal transition. The resistance of the fine-grained sample is lower than that of coarse-grained one below their respective $T_{MI}$s. This could be because of size driven transition of COI phase into FMM phase within a grain.[16] Beyond ~0.6 μm, enhanced influence of grain resistance might give rise to a drop in the $\rho_1/\rho_2$ ratio.

The $I_{th}$ – maximum current (indicated by arrows in the Fig. 3a top inset) which marks the onset of NDR regime – depicts a monotonic rise with the increase in grain size and $\rho_1/\rho_2$ ratio (Fig. 3a). This observation indicates that $I_{th}$ represents, most likely, the dielectric breakdown voltage of the grain boundaries. Observation of low $I_{th}$ for smaller grains and high $I_{th}$ for larger grains indicates that smaller grains offer smaller grain boundary resistance in spite of enhanced disorder at the grain-grain interface. Beyond a certain grain size (~0.6 μm), the $I_{th}$ crosses the current limit of our instrument and as a result we could not observe the onset of NDR in those cases.

Next, we turn to the central issue of our paper: the nature of variation of EER and EMR with grain size and grain boundary resistance (Figs. 3a,b main frames). EER decreases monotonically with increasing grain size across the entire range covered in this work. The EMR, on the other hand, follows a similar trend till a typical grain size of ~0.6 μm. Above this limit, EMR starts rising again. This correlation points out that below a grain size ~0.6 μm, one can observe colossal decrease in grain boundary resistance under simultaneous application of electric and magnetic fields. It is also noteworthy, in this context, that the temperature dependence of EMR is found to be very small for these fine-grained samples: $(1/EMR_{T=77 K}).(dEMR/dT) \sim 0.25$ over a temperature regime ~77 K-$T_{MI}$ whereas it varies within 0.6-1.50 for coarse-grained ones.

Thus, we observe a correlation between EER and EMR in the fine-grained LCMO systems. It is possible that the finer grain system comprises of well-dispersed *tiny* FMM islands[17] of the size of a few tens of nanometers (including nanoshorts at the grain boundaries) where the charge transport takes place primarily via SDS mechanism. The magnetoresistance versus field plot at a given temperature (bottom inset of Fig. 3b) for finer grained samples shows that there is no specific point of inflexion marking a sharp crossover of slope within the field range 0-10 kOe. This absence of inflexion also indicates that the charge transport across the grain boundaries for finer grained samples takes place via SDS mechanism. In this case, both the electric and magnetic field help in radically improving the connectivity across the grain boundaries through dielectric breakdown and enhancement of FMM phase volume fraction. As the grain size increases, the grain boundary resistance also increases while the grain-grain interface thickness decreases. The FMM islands join together to give rise to higher $T_{MI}$. Therefore, the impact of applied fields (current or magnetic) no longer remains as effective, leading to a drop in EER as well as EMR. Above a certain critical grain size, the charge transport mechanism, possibly, crosses over to SPT which gives rise to an enhancement in EMR as magnetic field improves the SPT in these half metals where majority and minority carrier spectra are split. The magnetic field helps in aligning the magnetic domains across the grain boundaries and hence maximizes the tunneling of majority carriers. The applied current, on the other hand, can neither trigger the onset of NDR in such a regime because of higher grain boundary resistance and consequent higher dielectric breakdown voltage nor can it influence the rotation/movement of the magnetic domains within the grains which is possible only in a multiferroic system.[18] Therefore, no correlation between EER and EMR is evident above a grain size ~0.6 μm.

## 4. Summary

In summary, we show that there is a similarity between the pattern of variation of EER and EMR with grain size (indicating a correlation between these two field driven effects) so long as the charge carrier transport across the grain boundaries follows SDS mechanism. For larger grains and higher grain boundary resistance, EMR could still be high as SPT mechanism governs the charge carrier transport across the grain boundaries, which is strongly influenced by the applied magnetic field. The EER, on the other hand, is small in this regime as electric field cannot influence the charge transport across the grain boundaries. This observation might help in devising a strategy for controlling the electrical resistance of such granular samples under simultaneous application of electric and magnetic fields. Extrapolation of the observed correlation between electro- and magnetoresistance indicates that samples with nanoscale grains (<200 nm) could offer even higher electro- and magnetoresistance under low field and hence be of tremendous potential for device applications.

## Acknowledgements

The authors PSD, AK, and DB acknowledge support under the CSIR network program CMM0022 during this work.

Table-I. List of few important parameters which characterize the fine-grained $La_{0.7}Ca_{0.3}MnO_3$

| Average Particle Size of Powder (nm) | Average Grain Size of Pellets (μm) | Lattice Parameters (Å) |
|---|---|---|
| 8.0 | 0.196 | a = 5.508; c = 13.339 |
| 8.5 | 0.340 | a = 5.436; c = 13.238 |
| 12.5 | 0.450 | a = 5.431; c = 13.397 |
| 13.7 | 0.560 | a = 5.457; c = 13.305 |
| 15.0 | 0.628 | a = 3.806 |
| 18.0 | 0.672 | a = 5.420; c = 13.198 |
| 80.0 | 1.000 | a = 5.521; c = 13.293 |

**Figure Captions:**

Fig. 1. (color online) (a) Room temperature x-ray diffraction patterns of nanoscale powder; * marked peak originates from sample holder; (b) and (c) SEM photographs of the grain morphology in sintered pellets of $La_{0.7}Ca_{0.3}MnO_3$. Insets: (b) and (c) the grain size distribution as estimated by the image analyzer Image-J.

Fig. 2. (color online) The variation of $T_{MI}$ (solid symbols) and $\rho_1/\rho_2$ (open symbols) with grain size for the fine-grained sintered $La_{0.7}Ca_{0.3}MnO_3$ samples. Dashed lines are guide to the eye. Inset: ρ-T plots for a few representative cases with arrows marking the $T_{MI}$.

Fig. 3. (color online) The variation of (a) electroresistance (solid symbols) and threshold current ($I_{th}$) (open symbols), and (b) extrinsic magnetoresistance with grain size for the fine-grained sintered $La_{0.7}Ca_{0.3}MnO_3$ samples. Dashed lines are guide to the eye. Insets: (a) current-voltage characteristics at ~77 K; $I_{th}$ is marked by arrow; (b) top: magnetoresistance versus normalized temperature and bottom: magnetoresistance versus field at ~87 K for a few representative cases.

Fig. 4. (color online) The resistance versus temperature data under zero and ~15 kOe magnetic field and the fitted lines are shown for a few representative samples with average grain size (a) 0.628 μm, (b) 0.56 μm, and (c) 0.45 μm.

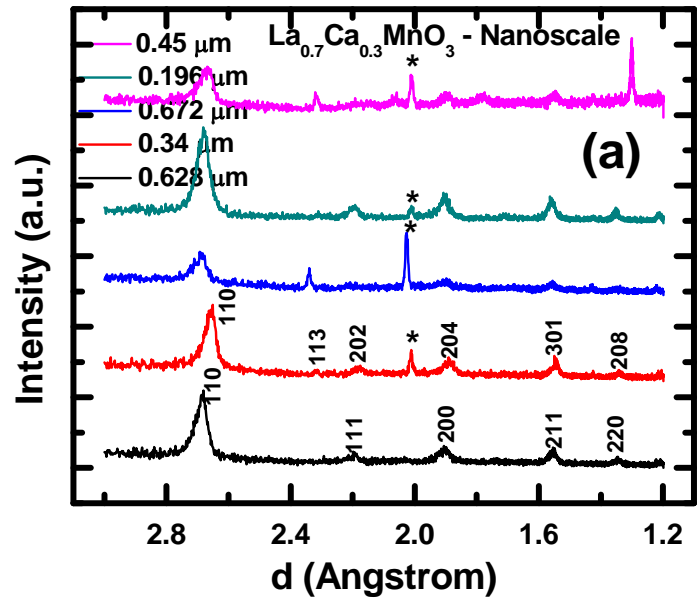
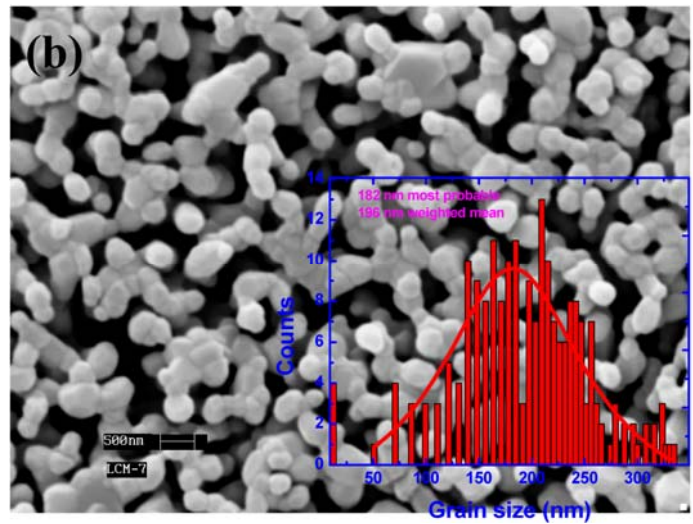
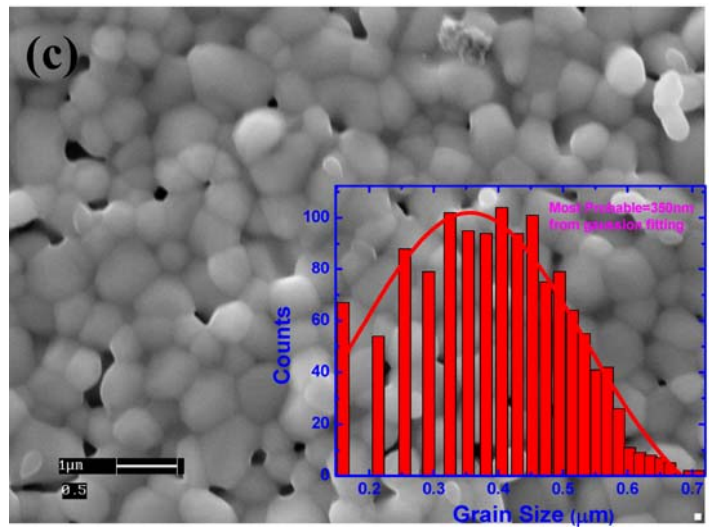

**Fig. 1.**

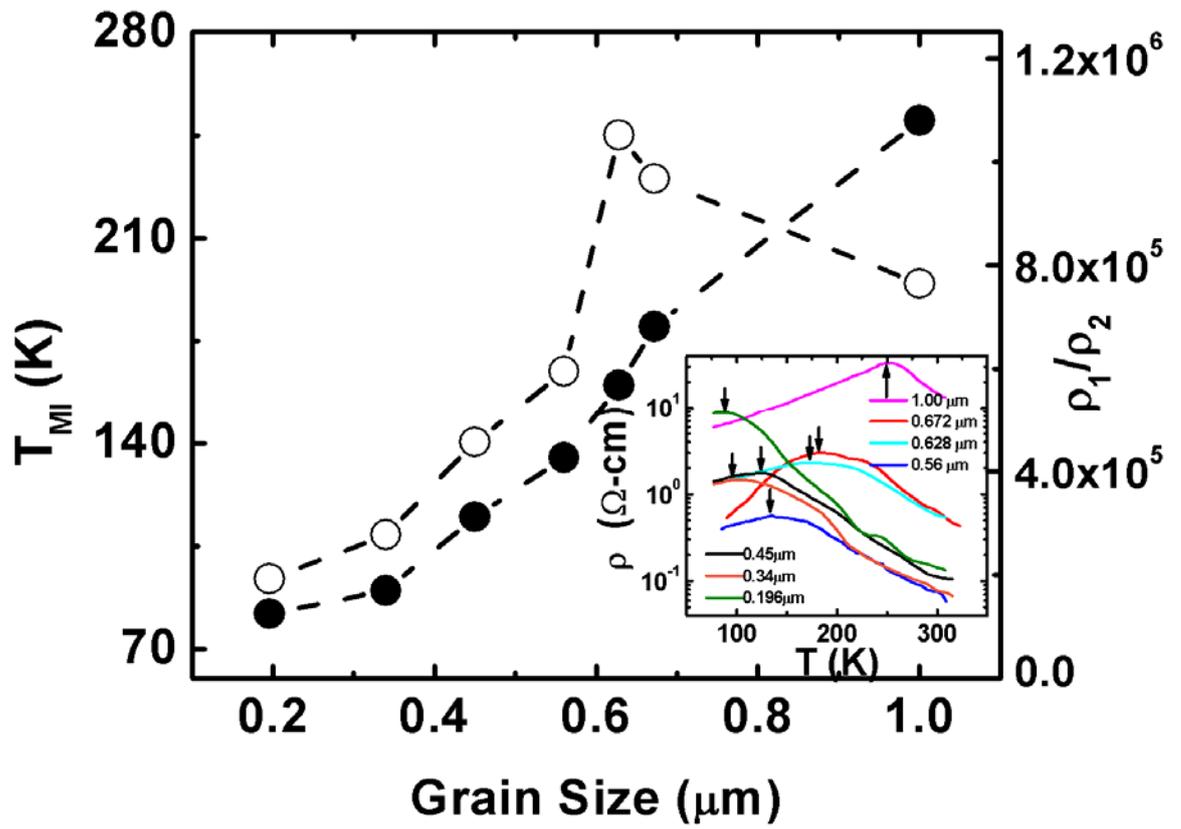

**Fig. 2.**

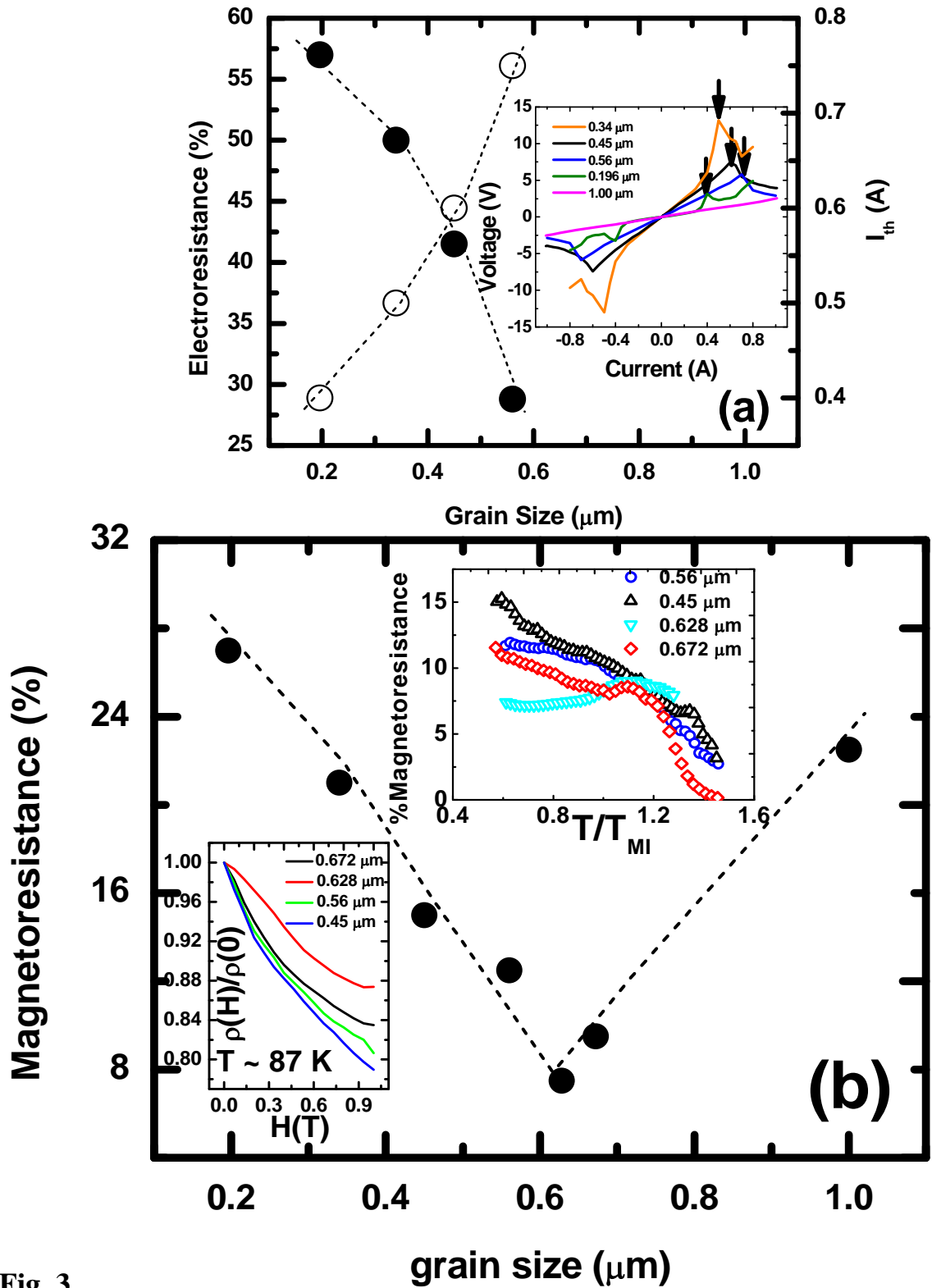

Fig. 3.

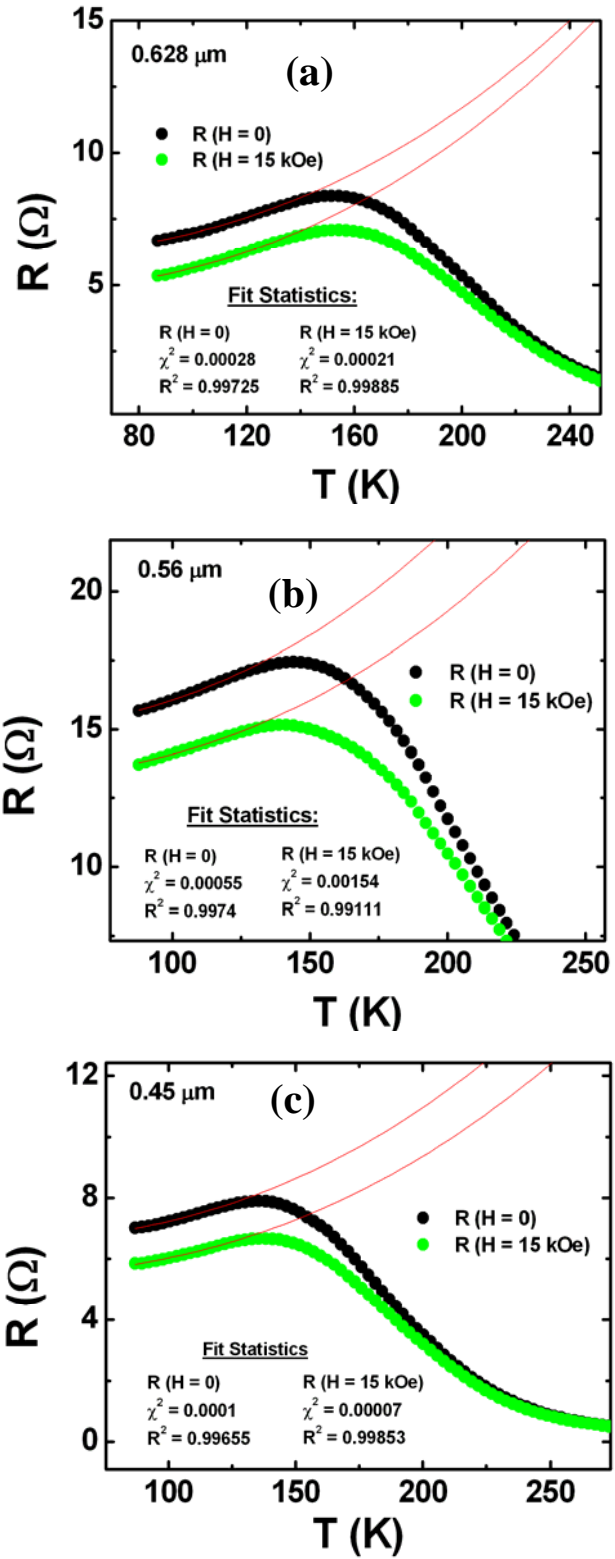

**Fig. 4.**